# 金刚石表面无定形碳氢薄膜生长的分子动力学模拟

张传国 [1], 杨勇 [1]*, 郝汀 [1], 张铭 [2]

[1](中国科学院固体物理研究所材料物理重点实验室，合肥 230031)
[2](北京工业大学材料科学与工程学院，北京 100124)

**摘要**：利用分子动力学模拟方法研究了 $CH_2$ 基团轰击金刚石(111)面所形成的无定形碳氢薄膜 (a-C:H) 的生长过程。结构分析表明，得到的无定形碳氢薄膜中碳原子的局域结构(如 C-C 第一近邻数)与其中氢原子的含量密切相关。$CH_2$ 基团入射能量的增加会导致得到的薄膜的氢含量降低，从而改变薄膜中类 $sp^3$ 成键碳原子的比例。

**关键词**：无定形碳氢薄膜，金刚石表面，分子动力学模拟

**PACS:** 81.15.Aa, 81.05.uj

---

*通讯作者. E-mail: yyang@theory.issp.ac.cn

# 1. 引言

无定形碳氢薄膜（以下简写为 a-C:H），由于其具有较为优异的物理与化学性能，如高硬度、抗耐磨、低摩擦系数、光学透光性及化学惰性等特性，因而广泛应用于磁存储器件中的保护涂层、汽车工业中的耐磨涂层、以及光学涂层、各种医学领域中的生体相容涂层、惯性约束反应堆靶丸制造等领域 [1-6]。决定 a-C:H 薄膜性能的关键，在于组成的非晶态和微晶态结构中碳-碳原子间的 $sp^2$ 与 $sp^3$ 杂化键的相对比例。如果制备出的无定形碳膜中 $sp^3$ 杂化占优，局域成键结构类似于金刚石，即类金刚石相比例含量居多，因此通常称为类金刚石 (Diamond Like Carbon，简写为DLC) 薄膜，其强度及摩擦学性能较优；如果 $sp^2$ 杂化占优则类石墨相的含量居多，将导致薄膜的上述性能的劣化。最近在美国国家点火装置 (NIF) 进行的惯性约束核聚变首次获得增益大于 1 的能量输出 [6]，采用的靶丸材料就是掺杂少量 Si 的 DLC 碳氢薄膜。采用常用的实验手段如等离子体增强化学气相沉积法 (PECVD)，磁控溅射等实验手段，通过改变沉积工艺参数，可以制备出的不同结构与性能的类金刚石膜、类石墨膜以及类富勒烯膜的 a-C:H 薄膜 [7-9]。这表明了 a-C:H 薄膜的结构形态与性能容易受到外界形成条件的影响。为了优化 a-C:H 薄膜的性能，如何确定薄膜制备工艺条件与薄膜微观结构和性能之间的对应关系、深入理解薄膜微观结构形成过程中原子之间的成键方式对制备出结构稳定、性能优异的 a-C:H 薄膜具有相当重要的意义。<u>实验上，人们还研究了通过掺杂其他的无机材料 (如 ZnO 颗粒，Ti、Al、N 等元素) 对 DLC 薄膜的结构、形貌以及力学性质的影响 [10-12]。</u>但是，现有的实验手段尚不能从原子层面评价薄膜的三维原子结构形态，这就大大增加了人们从微观机理上认识薄膜结构形成过程和后续性能之间关系的难度。因此，计算机模拟被认为是一种理想的研究 a-C:H 薄膜的沉积过程和薄膜性能的有力工具。

基于经验势的分子动力学（MD）模拟由于可以研究大规模原子的动态过程，已经被广泛用在 a-C:H 薄膜的生长过程研究中 [13-15]。马天宝等人使用 MD 模拟方法研究了 2-3 nm 厚的 DLC 薄膜在金刚石基体(100)面上的生长过程，发现稳定生长的中间区域的存在对保证 DLC 薄膜的性能至关重要；随着入射原子能量的增加，$sp^2$ 与 $sp^3$ 杂化比例和薄膜密度也增加并且达到一个饱和值；当入射能量增加至 20-60 eV 时，能获得最优的薄膜结构特性 [16]。为了模拟 PEVCD 薄膜生长过程中的多种类入射原子（如 C、H、$C_2$、$C_2H_2$、$C_2H_3$、$C_2H_4$、$C_2H_5$ 等）对 a-C:H 薄膜结构形成过程的影响，Neyts 等人系统地研究了带有低入射能的 C、H、$C_2$、$C_2H$、圆环型 $C_3H$ 和直线型 $C_3H$ 混合源对薄膜生成结构的影响，发现碳原子的 $sp^3$ 杂化比例随着薄膜中 H 原子的比例增加而增加，并且薄膜中 H 含量与入射混合源中 H 原子含量之间呈近似线性依赖关系 [17-20]。然而他们的 MD 模拟工作中采用了入射能较低（平均能量仅为 0.13 eV/原子团）的混合原子团来展开研究，导致生成的薄膜密度和 $sp^3$ 杂化比例较低，并且由于入射原子团的协同效应致使难以区分各个入射原子团对薄膜结构性能的影响机制。为此，Quan 等人分别使用入射能量不同的单原子/原子团（如 40 eV 的 C，20 eV 的 H，3.25 eV、6.5 eV、13 eV、26 eV、39 eV、65 eV、97.5 eV 以及 130 eV 的 CH）研究了轰击生成的 a-C:H 薄膜在金刚石表面的生长过程，发现几乎所有的 H 原子都与薄膜中的碳原子形成化学键，并且 $sp^3$-C 比例随入射 H 原子增加而增加 [7, 21, 22]。

本文使用 MD 模拟研究 $CH_2$ 原子基团轰击金刚石表面过程中 a-C:H 薄膜的沉积生长过程，通过研究 $sp^2$ 与 $sp^3$ 杂化比例、薄膜密度以及径向分布函数等研究薄膜的结构，并与 CH、$C_2H$ 等轰击行为结果对比，考察能量不同的入射 $CH_2$ 原子基团在薄膜生长中扮演的角色，为制备出 $sp^3$ 杂化占优的光滑致密 a-C:H 薄

膜提供理论支持。

## 2. 模型及其模拟方法

MD 模拟采用 LAMMPS 程序包 [23] 完成。利用 Stuart 等人发展的自适应分子间反应经验键级 (airebo) 势函数来描述 C-C、C-H 原子之间的相互作用 [24]，截断距离取 3.0 Å。模拟采用的模型如图 1(a) 所示，$CH_2$ 基团从(111)方向，即 Z 方向入射。金刚石(111)取向的衬底由众多厚度为 ~0.52 Å 的双原子层堆积而成，每个双原子层的间隔为~ 2.06 Å，见图 1(b)。模拟的超原胞包含 18 个双原子层和一个单原子层。其中，单原子层及其沿着 Z 方向最近邻的一个双原子层的位置固定，用来模拟金刚石的体材料的性质。沿 XY 方向的表面超原胞的尺寸为 $4\sqrt{2}a_0 \times 2\sqrt{6}a_0$，$a_0$ ~ 3.57 Å，为晶格常数。沿 Z 方向的超原胞的长度为$14\sqrt{3}a_0$，其中金刚石原子层的高度为$6\sqrt{3}a_0$，真空层厚度为$8\sqrt{3}a_0$，用以模拟表面。基体部分由 2368 个碳原子组成，整个体系在 XYZ 方向采用周期边界条件。靠近底部固定层的 3 个双原子层 (厚度为$\sqrt{3}a_0$) 通过速度再标度的方法 (velocity re-scale) 耦合在温度为 523 K 的热浴中，这是热浴原子层，用以保持系统的温度。更上层的原子层为自由弛豫层。入射原子团与表面发生作用时产生的热量会通过自由弛豫层向热浴原子层耗散。在模拟薄膜生长之前，衬底已经在温度为 523 K 的环境中采用正则系综 (NVT) 系综弛豫 10 ps。在模拟薄膜生长过程中，模拟体系采用微正则 (NVE) 系综。

入射团簇以碳原子位置标记入射团簇的位置。在 XY 方向上随机选择入射位置，Z 方向上入射位置为离表面高度为$5\sqrt{3}a_0$。这样足以保证刚入射的团簇不会受到表面的作用。至于入射团簇 C-H 键的取向，我们采用随机取向 (团簇绕 C

原子刚体转动)，所以在这里没有考虑团簇原子化学键取向的影响。所有团簇的入射方向都垂直于表面。入射基团的动能（$E_k = \frac{1}{2}mv^2$）分别为 20 eV 和 100 eV。模拟的时间步长依赖原子的速度或者所受的力的大小，是可变的，变化范围为 0.0001-0.001 ps，这样可以提高计算效率和精度。$CH_2$ 基团每隔 1 ps 入射一次，这个时间已经足够表面弛豫达到热平衡。需要指出的是，由于模拟的时间步长变化，所以对入射能量不同的基团，模拟相同的步数，对应的模拟时间上会有一些差别。例如，模拟 $10^6$ 步之后，入射能量为 20 eV 的体系对应的时间大约 382 ps，而入射能量为 100 eV 的体系对应约 445 ps。

## 3. 计算结果及讨论

图 2 显示了在分别模拟 $10^5$、$5×10^5$ 以及 $10^6$ 步 (对应的模拟时间分别为 $t_1$ = 39.459 ps, $t_2$ = 192.562 ps, $t_3$ = 382.056 ps) 之后，能量为 20 eV 的 $CH_2$ 基团轰击金刚石(111)面得到的 a-C:H 薄膜的侧视图。很明显，随着模拟时间的增加，在金刚石表面上入射并沉积下来的碳原子和氢原子的量越来越大，逐渐形成无定形的碳氢薄膜。图 3 显示了在分别模拟 $10^5$、$5×10^5$ 以及 $10^6$ 步 (对应的模拟时间分别为 $t_1` $ = 45.041 ps, $t_2` $ = 226.023 ps, $t_3` $ = 445.232 ps) 之后，能量为 100 eV 的 $CH_2$ 基团轰击金刚石(111)面得到的 a-C:H 薄膜的形貌图。类似的，碳原子和氢原子在基底上的沉积量随着模拟时间的增加而增大。另一方面，图 2 和图 3 相比较，一个显著的不同之处是，对于相近的时间尺度，图 3 中氢原子相对于碳原子的含量明显下降。

我们进一步计算了在模拟过程中，沉积在基底的氢原子在 a-C:H 薄膜中的原子含量比例随时间的变化，计算结果如图 4 所示。在模拟开始的时候 (t = 0),

CH$_2$基团刚刚入射，所以氢原子的含量比例自然就是 2/(2+1) = 2/3 ~ 0.667。对入射能量为 20 eV 的基团，当模拟时间 t < 200 ps 时，尽管有些涨落，氢原子在薄膜中的含量总体呈现出上升趋势。在模拟时间超过 200 ps 之后，氢原子的含量比例趋于一个稳定值: ~ 0.4。对于入射能量为 100 eV 的基团，当模拟时间超过 150 ps 之后，碳氢薄膜中氢原子的含量趋于稳定，比例大约是 0.09。因此，对模拟所允许的时间尺度来说，入射能量为 20 eV 的 CH$_2$基团轰击金刚石(111)面获得的无定形碳氢薄膜中氢的含量大约是 100 eV 的 CH$_2$基团轰击获得碳氢薄膜中氢含量的 4 倍。对上述两种情形，得到的碳氢薄膜中氢的含量都小于 0.667，其原因是: 高速入射的 CH$_2$基团和金刚石表面碰撞之后，碳氢键断裂，其余的一部分能量转化为分离后的入射原子的反射动能。离解后的碳原子和表面以及邻近原子可以形成 4 个键 (C-C 单键能量 ~ 346 kJ/mol [25])，而氢原子至多和表面形成 1 个键 (C-H 键能量 ~ 411 kJ/mol [25])。更强的相互作用使得碳原子被表面俘获的概率更大。另外，从薄膜可能的微观构造来看，由于氢原子只能和近邻原子形成 1 个稳定化学键 (C-H 或者 H-H)，而碳原子则可以形成 4 个 (C-C，C-H)。稳定的薄膜结构必然是碳原子数量占优。被表面反射的碳、氢原子如果在势场的有效作用距离之外，则以恒定的速度离开模拟体系 (从体系里移除)。因此，我们的模拟体系事实上由一系列 NEV 系综构成的 (即对每一组时间上连续的、粒子数相同的构型，模拟的是 NEV 系综)。

为了深入研究入射能量造成的差别，我们计算了能量分别为 20 eV 和 100 eV 的 CH$_2$基团轰击之后，金刚石基底以及得到的碳氢薄膜沿着 Z 方向的密度变化，结果见图 5: 左版面对应能量为 20 eV 的情形，右版面对应能量为 100 eV 的情形。对入射能量为 20 eV 的基团，随着模拟时间的增加，当 t > t$_1$ 时，在 Z 方向的高

度为 Z ~ 35 Å 处，体系的密度开始低于金刚石体相的密度 (~ 3.51 g/cm$^3$)。这表明入射基团对最上面的碳原子层的溅射造成了结构缺损。从 ~ 35 Å 到 ~ 40 Å 左右的薄膜区域，体系的密度迅速地下降。密度为 0 的区域对应于真空层。显而易见的是，随着模拟时间的增加 ($t_1 \rightarrow t_2 \rightarrow t_3$)，碳氢薄膜的厚度不断增加，真空层厚度相应减少。对入射能量为 100 eV 的基团，当 t = $t_3$` 时，高度 Z ~ 30 Å 处，体系的密度开始低于金刚石的密度。这是可以理解的，因为入射粒子能量的增加会增大其对基底的穿透和溅射能力。与 20 eV 入射的情形不同，薄膜的密度并非单调下降：在高度 Z ~ 40 Å 处，出现一个峰位，对应的密度大于 20 eV 的情形。这清楚的表明了入射能量不同所得到的碳氢薄膜在微观结构上的差别。

我们进一步计算了轰击生成的碳氢薄膜的 C-C、C-H 原子的径向分布函数 $g_{CC}$ 和 $g_{CH}$。我们着重考察模拟时间在 150 ps 之后的情形。对 20 eV 入射的情形，我们考察两个时间点的薄膜构型：$t_2$ = 192.562 ps, $t_3$ = 382.056 ps。对 100 eV 入射的情形，则考察 $t_2$` = 226.023 ps 和 $t_3$` = 445.232 ps 对应的位型。计算的结果如图 6 所示。对于 $g_{CC}$ 和 $g_{CH}$，都出现了第一和第二个峰位，表明得到的薄膜是短程有序而长程无序的。能量为 20 eV 和 100 eV 这两种情形，分别考察的两个时间点对应的径向分布函数 $g_{CC}$ 的第一个峰位 (C-C 第一近邻) 都处于 ~ 1.40 Å，前者的第二个峰位 (C-C 第二近邻) 位于 ~ 2.49 Å，而后者的第二近邻位于 ~ 2.43 Å。这表明后者的碳原子排列更加紧密。与此同时，$g_{CC}$ 第一个峰位的展宽都处于 1.17 Å 至 1.75 Å 之间。这对应着无定形碳氢薄膜中 C-C 键的键长取值范围，包括了石墨中 C-C 键长 (~ 1.42 Å)，以及金刚石中的 C-C 键长 (~ 1.54 Å)。对 $CH_2$ 入射能量为 20 eV 和 100 eV 这两种情形，径向分布函数 $g_{CH}$ 的第一个峰位分别处于 ~ 1.06 Å 和 1.05 Å，这和典型的 C-H 键长取值 (~ 1.1 Å) 吻合。前者第一个峰位的

展宽处于 ~0.9 Å 至 ~1.36 Å 之间，而后者则处于 ~0.9 Å 至 ~1.30 Å 之间。对 20 eV 以及 100 eV 入射的情形，$g_{CH}$ 的第二个峰位分别处于 ~2.13 Å 和 ~2.15 Å。当 C-C 以及 C-H 的距离处在第二个峰位之后，薄膜的结构处于无序的状态。不同的模拟时间点得到的薄膜的局域结构 ($g_{CC}$ 和 $g_{CH}$) 是类似的。这表明，在模拟 150 ps 之后，薄膜开始成形。这和前面图 4 得到的氢含量在 t > 150 ps 趋于稳定的结果是一致的。

前面提到，碳氢薄膜中成键为 $sp^2$ 和 $sp^3$ 的碳原子的数量会对薄膜的宏观性质如力学韧性以及光洁度有着重要的影响。我们分别对模拟 $10^6$ 步的过程中，上述两种情形 (20 eV 和 100 eV 入射) 的碳原子的 C-C 配位数 (Coordination Number，以下简写为 CN) 随时间的演化做了计算。碳氢薄膜中配位数 CN = 2, 3, 4 的碳原子的数量随时间的变化关系如图 7 所示。显而易见的是，随着模拟时间的增加以及碳原子总沉积量的增大，配位数 CN = 2, 3, 4 的碳原子的绝对数量也在增长。对 20 eV 入射的情形，较为有趣的是，配位数 CN = 2 和 3 的碳原子数量在整个模拟过程中保持大致相等。对 100 eV 入射的情形，CN = 3 的碳原子数量最多，超过了 CN = 2 和 CN = 4 的碳原子之和。对两种情形，配位数 CN = 4，也就是 $sp^3$ 成键或接近于 $sp^3$ 成键的碳原子的数量最少。就绝对数量而言，20 eV 入射产生的薄膜的 CN = 4 的碳原子数量大于 100 eV 入射的情形。

在各模拟 $10^6$ 步之后，20 eV 的 $CH_2$ 入射产生薄膜的碳原子和氢原子数量分别是 213 和 140，而 100 eV 的 $CH_2$ 入射产生薄膜的碳、氢原子数量则分别是 331 和 30。能量为 20 eV 的 $CH_2$ 入射获得的薄膜中的氢原子不论是相对含量还是绝对数量都明显大于 100 eV 入射的情形。氢原子含量的增加会使得碳氢薄膜中 $sp^3$ 成键 (CN = 4) 的碳原子的数量增加，这个结果和先前用 CH 以及 $C_2H$ 基团轰击

金刚石(111)面 [7, 17] 生成的碳氢薄膜的研究结果是类似的。氢原子的存在，使得薄膜的局域结构多孔化，这客观上降低了薄膜的局域密度 (见图 5, Z ~ 40 Å 处的密度)，抑制了 C-C 键长更短的 $sp^2$ 的成键概率，使得 $sp^3$ 的成键机会得以增加。

那么，为什么入射能量更大的 $CH_2$ 基团轰击获得薄膜氢含量更低呢？其原因主要是，$CH_2$ 基团入射到金刚石表面上，其动能主要耗散在以下几个方面：1. 传递给表层原子致使其散射、溅射；2. 传递给已经沉积在基底上的初步形成的碳氢薄膜的碳、氢原子，其散射、溅射；3. $CH_2$ 自身 C-H 键的断裂以及散射。入射 $CH_2$ 基团的能量越高，$CH_2$ 自身 C-H 键断裂得越彻底，轰击之后背向反射出来的氢原子或者 CH 基团的能量相应的更高，被表面或者已沉积下来的碳氢薄膜俘获的概率也就越小，自然造成了生成的薄膜中氢含量下降。

## 4. 结论

通过分子动力学 (MD) 模拟，我们研究了能量不同 (20 eV 和 100 eV) 的 $CH_2$ 基团轰击金刚石(111)面形成无定形碳氢薄膜的动态过程。对 C-C 和 C-H 径向分布函数的分析表明，得到的碳氢薄膜是局域有序的。对薄膜的密度、氢含量以及 C-C 配位数为 2，3，4 的碳原子数量的分析指出，碳氢薄膜中氢的含量会直接影响到其中 $sp^3$ 成键 (C-C 配位数 = 4) 的碳原子的数量。在一定范围内，薄膜中氢的相对含量高的薄膜的 $sp^3$ 成键的碳原子数量更多一些。而轰击生成的薄膜中氢含量和 $CH_2$ 基团的入射能量密切相关。在我们研究的能量范围内，更高的入射基团能量伴随着更低的氢含量。在将来的研究中，我们还将考虑入射基团的 C-H 键取向，基底表面结构以及温度对薄膜生长的影响。

# Molecular dynamics simulations of the growth of thin amorphous hydrogenated carbon films on diamond surface

ZHANG Chuan-guo[1], YANG Yong[1]*, HAO Ting[1], ZHANG Ming[2]

[1](Key Laboratory of Materials Physics, Institute of Solid State Physics, Chinese Academy of Sciences, Hefei 230031, China)
[2](The College of Materials Science and Engineering, Beijing University of Technology, Beijing 100022, China)

**Abstract:** The growth of thin amorphous hydrogenated carbon films (a-C:H) on diamond (111) surface from the bombardment of $CH_2$ radicals is studied using molecular dynamics simulations. The structural analysis shows that the local structure (e.g., the first coordination number of C atoms) of a-C:H depends critically on the content of hydrogen. The increase of kinetic energy of incident radicals leads to the decrease of hydrogen content, which subsequently changes the ratio of $sp^3$ bonded C atoms in a-C:H.

**Keywords:** amorphous hydrogenated carbon films, diamond (111) surface, molecular dynamics simulations

**PACS:** 81.15.Aa, 81.05.uj

---------------------------------------------------------------------------------------------------------------

*Corresponding Author. E-mail: yyang@theory.issp.ac.cn

**图形及文字说明**

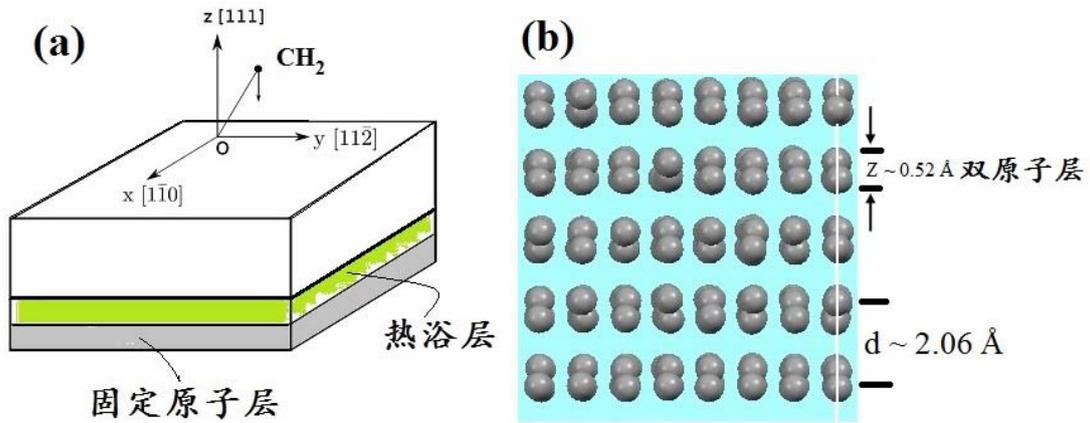

图 1 (a) 金刚石(111)面及入射基团示意图；(b) 金刚石沿(111)方向的双原子层示意图。

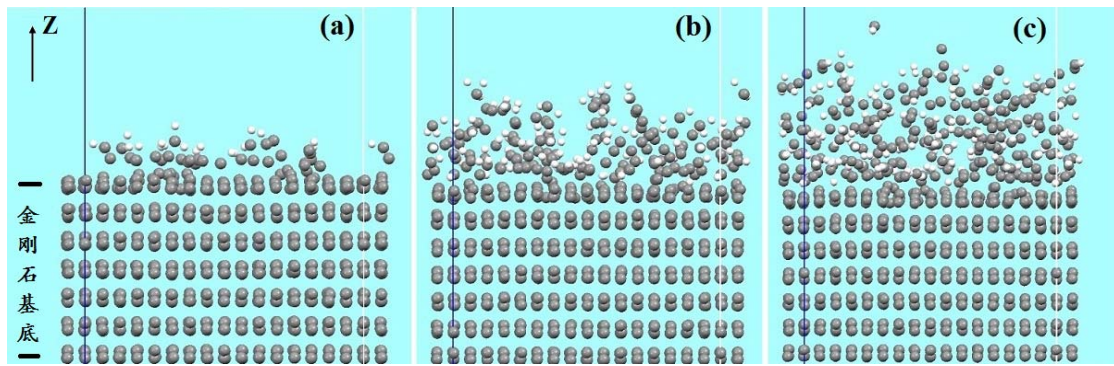

图 2 能量为 20 eV 的 $CH_2$ 轰击金刚石表面得到的碳氢薄膜侧视图 (a) 时刻 $t_1$ = 39.459 ps; (b) 时刻 $t_2$ = 192.562 ps; (c) 时刻 $t_3$ = 382.056 ps。图中灰色原子表示碳，白色原子表示氢。模拟的超原胞沿 X 方向的左右边界用蓝线和白线标出。

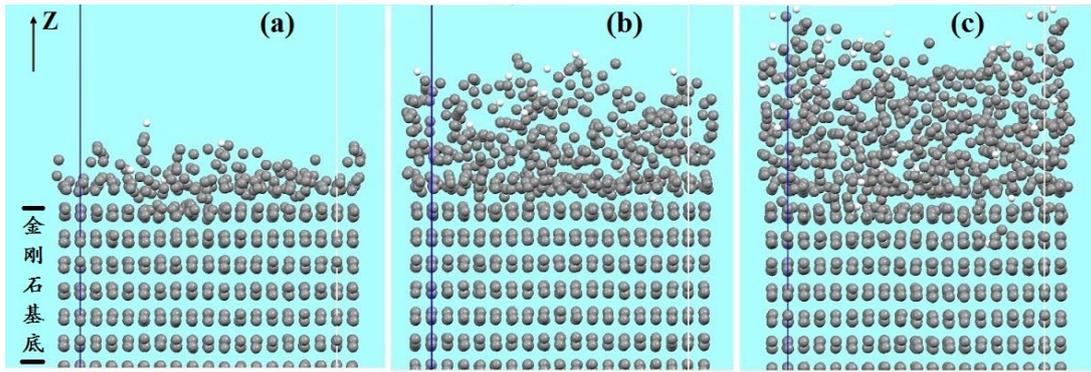

图3 能量为100 eV的$CH_2$轰击金刚石表面得到的碳氢薄膜侧视图 (a) 时刻 $t_1` =$ 45.041 ps; (b) 时刻 $t_2` = 226.023$ ps; (c) 时刻 $t_3` = 445.232$ ps。 图中灰色原子表示碳，白色原子表示氢。模拟的超原胞沿X方向的左右边界用蓝线和白线标出。

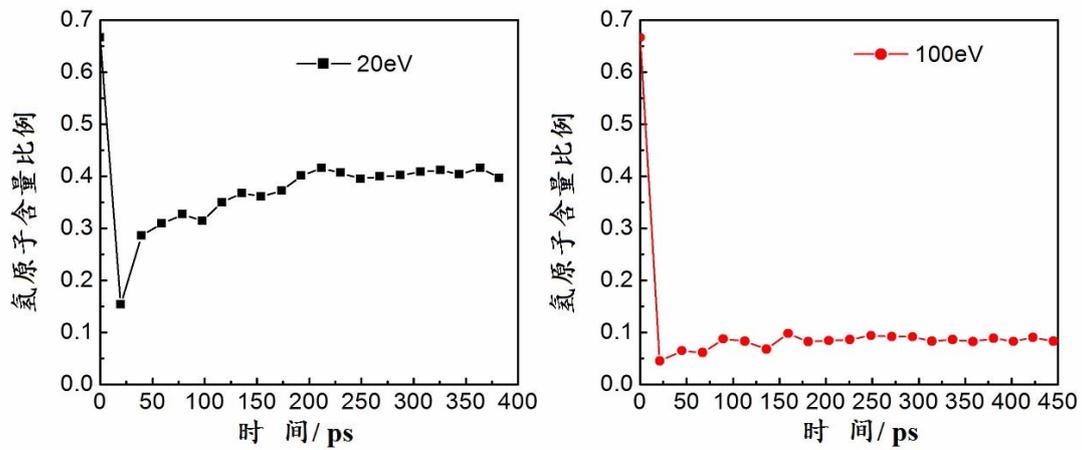

图4 能量为20 eV (左版面) 和100 eV (右版面) 的$CH_2$基团轰击金刚石表面得到的碳氢薄膜中氢的含量比例随时间的演化示意图。

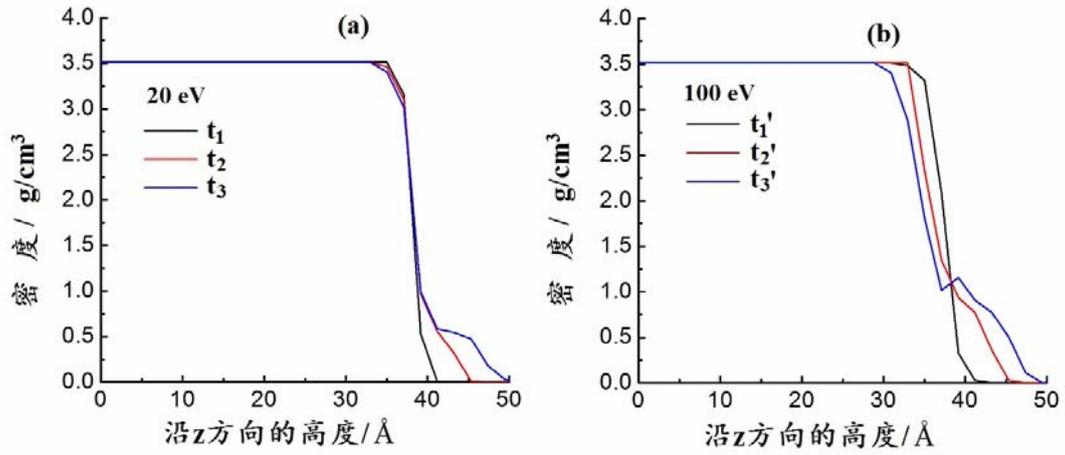

图 5 金刚石基底以及碳氢薄膜的沿 Z 方向在不同时刻的密度变化: (a) 20 eV 的 $CH_2$ 基团轰击；(b) 100 eV 的 $CH_2$ 基团轰击。各时刻 ($t_1, t_2, t_3, t_1`, t_2`, t_3`$) 的含义与前面相同。

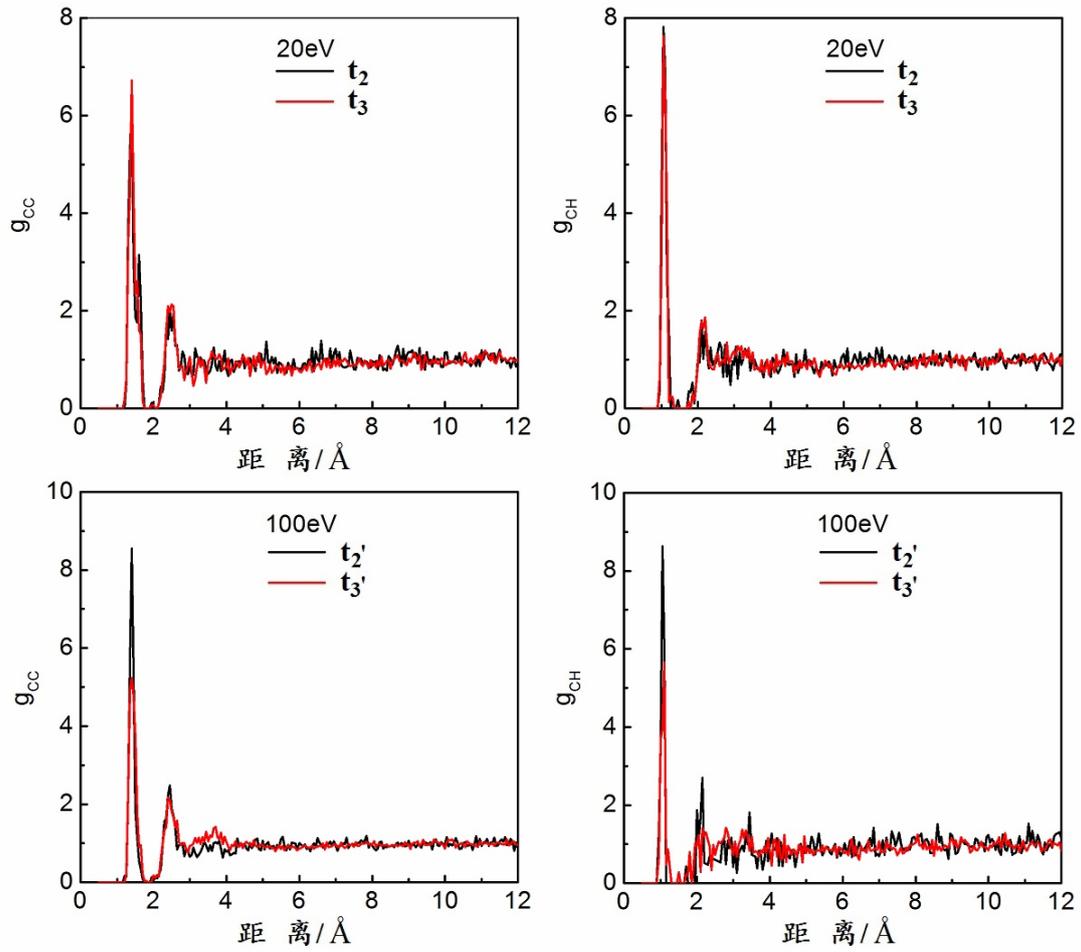

图 6 在不同的时刻，能量为 20 eV 和 100 eV 的 CH$_2$ 基团轰击金刚石基底生成的碳氢薄膜的 C-C、C-H 径向分布函数。

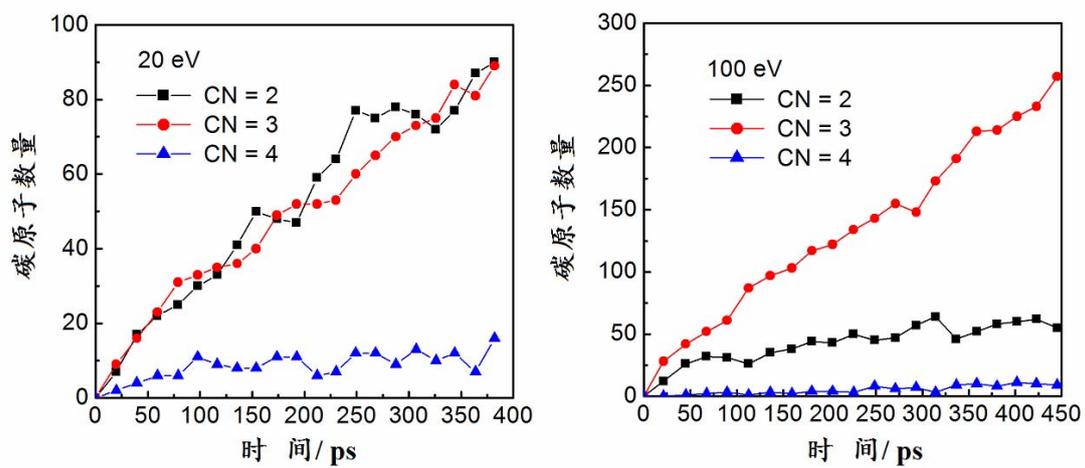

图 **7** 能量为 20 eV 和 100 eV 的 $CH_2$ 基团轰击生成的碳氢薄膜中 C-C 配位数 CN = 2, 3, 4 的碳原子数随时间的变化关系图。